\documentclass[a4paper, 11pt]{article}
\usepackage{jheppub}

\title{Towards $Wb\overline b + j$ at NLO with an automatized approach
to one-loop computations}

\author[a,b]{Laura Reina,}
\author[a]{Thomas Schutzmeier}

\affiliation[a]{Physics Department, Florida State University \\
Tallahassee, FL, 32306-4350, USA}

\emailAdd{reina@hep.fsu.edu}
\emailAdd{tschutzmeier@hep.fsu.edu}

\abstract{We present results for the $\mathcal O(\alpha_s)$ virtual
  corrections to $qg \rightarrow W b \overline b q'$ obtained with a
  new automatized approach to the evaluation of one-loop amplitudes in
  terms of Feynman diagrams. Together with the $\mathcal O(\alpha_s)$
  corrections to $q \overline q' \rightarrow W b \overline b g$, which
  can be obtained from our results by crossing symmetry, this
  represents the bulk of the next-to-leading order virtual QCD
  corrections to $W b \overline b + j$ and $W b + j$ hadronic
  production, calculated in a fixed-flavor scheme with four light
  flavors. Furthermore, these corrections represent a well defined and
  independent subset of the 1-loop amplitudes needed for the NNLO
  calculation of $Wb\bar{b}$. Our approach was tested against several
  existing results for NLO amplitudes including selected $\mathcal
  O(\alpha_s)$ one-loop corrections to $W + 3\,j$ hadronic production.
  We discuss the efficiency of our method both with respect to
  evaluation time and numerical stability.}

\newcommand{\Wbbd}{d g \rightarrow W b \overline b u}
\newcommand{\Wddg}{\overline u d \rightarrow W d \overline d g}

\begin{document}

\maketitle

\section{Introduction} 
\label{sec:intro}

One of the most important background processes to single-top
production as well as searches for a light Higgs boson is the
associated production of a $W$~boson with a pair of massive bottom
quarks, contributing to both the $W + b\text{-jet}$ and $W +
2\,b\text{-jet}$ signatures. The precise theoretical knowledge of
these processes provides moreover an excellent probe of our current
understanding of QCD when confronted with measurements in various
kinematic regimes at high-energy hadron colliders.

At the Tevatron $p\overline p$ collider at Fermilab, the
cross sections for $W$ boson + $b$~jets production have been measured
in different forms by both the CDF~\cite{Aaltonen:2009qi} and
D0~\cite{Abazov:2004jy} collaborations and the Large Hadron $pp$
Collider (LHC) at CERN is expected to provide additional experimental
insight with increased precision. A first measurement from the ATLAS
collaboration has recently been published~\cite{Aad:2011kp} and new
measurements with better statistics are expected by the end of the
year.

On the theory side, next-to-leading order (NLO) QCD corrections to $W$
production with up to two jets containing at most one heavy $b$ jet
are known \cite{Campbell:2006cu} and cross sections for $W +
2\,b\text{-jets}$ were determined both in the massless $b$-quark
approximation \cite{Bern:1997sc,Bern:1996ka, Ellis:1998fv,
  Campbell:2002tg, Campbell:2003hd} and including $b$-quark mass
effects \cite{Febres Cordero:2006sj, Cordero:2007ce, Cordero:2009kv,
  Badger:2010mg} at the same level of precision.  From existing NLO
QCD calculations, the theoretical prediction for the production of
$W+2$ jets with at least one $b$ jet has been provided in
Ref.~\cite{Campbell:2008hh} and compared to the
CDF~\cite{Aaltonen:2009qi} and ATLAS~\cite{Aad:2011kp} measurements in
Refs.~\cite{Cordero:2010qn} and \cite{Caola:2011pz} respectively.
Recently, the NLO calculation of $Wb\overline b$ has been interfaced
with shower Monte Carlo generators using both \texttt{POWHEG}
\cite{Oleari:2011ey} and \texttt{MC@NLO}~\cite{Frederix:2011qg}. The
study of the NLO corrections to $Wb\bar{b}$ shows, however, that the
NLO theoretical prediction for $Wb\overline b$ production is plagued
by large renormalization and factorization scale uncertainties in
particular at the LHC~\cite{Cordero:2009kv}. In fact, at this order of
perturbative QCD, a new $qg$ initiated channel with an additional
parton in the final state ($qg \rightarrow Wb \overline b q'$) opens
up and, being a tree level process, introduces a strong scale
dependence.  This effect is particularly pronounced at the LHC, where
the NLO $qg$ channel competes with the $q\overline q^\prime$ channel
due to the substantial initial state gluon density.  Only a complete
NNLO calculation of $pp(p\bar{p}) \rightarrow Wb\overline b$ can be
expected to reduce this spurious scale dependence and give a
theoretical prediction consistent at this order of QCD.  However, this
requires the evaluation of two-loop virtual corrections to a massive
$2 \rightarrow 3$ process as well as one-loop corrections to $2\to 4$
massive processes, and single and double particle emissions through
real corrections: a truly difficult task.  In this paper we focus on
and present results for one of the many contributions to
$pp(p\bar{p})\rightarrow Wb\overline b$ at NNLO: the $\mathcal
O(\alpha_s)$ virtual corrections to the $qg\rightarrow W b \overline b
q'$ channel, keeping the full bottom-quark mass dependence. Results
for the $\bar{q}g\rightarrow W b \overline b \bar{q}'$ channel are
identical at the partonic level and in the following it will be
understood that $qg\rightarrow Wb\bar{b}q^\prime$ refers to both
channels.

Our choice is motivated by the following considerations. First of all,
the $\mathcal O(\alpha_s)$ virtual corrections to the $qg \rightarrow
W b \overline b q'$ channel are a well-defined independent piece of
the overall NNLO calculation of $Wb\bar{b}$ hadroproduction. It can be
directly translated into the analogous $\mathcal O(\alpha_s)$ virtual
corrections to $q \overline q' \rightarrow Wb\overline b g$ by partial
crossing of initial and final states.  Once interfered with the
corresponding tree level amplitudes, they provide a self-standing and
well-defined part of the one-loop contributions to the full NNLO
$Wb\bar{b}$ cross section, namely the one-loop virtual contributions
from $2\rightarrow 4$ processes. The remaining one-loop corrections
come from the interference of the one-loop amplitude for the
$2\rightarrow 3$ process ($q\bar{q}^\prime\rightarrow Wb\bar{b}$) with
itself, and is not part of this study.  Furthermore, when complemented
with the corresponding real corrections to $qg \rightarrow W b
\overline b q'$ and $q \overline q' \rightarrow Wb\overline b g$, our
calculation completely determine the NLO cross sections for both $W b
\overline b + j$ and, within a fully consistent four-flavor-number
scheme, $W b + j$ production, i.e. for the production of a $W$ boson
with one or two $b$ jets plus a light jet, where the difference
between the two processes is just the number of $b$ jets
tagged in the final state (the parton level processes being the same
in the four-flavor-number scheme).  Since NLO real-emission
contributions nowadays can be determined in a mostly automatized
fashion with the help of existing packages, for instance
\texttt{SHERPA} \cite{Gleisberg:2008ta}, as well as NLO Monte Carlo
frameworks as \texttt{POWHEG}~\cite{Nason:2004rx,Frixione:2007vw} and
\texttt{MC@NLO}~\cite{Frixione:2002ik}, the virtual one-loop
corrections that we present in this paper constitute the only missing
piece for the NLO QCD cross section prediction of the above processes
and are therefore highly desirable. Both in this and the previous
case, our calculation should contribute to reduce the theoretical
uncertainty from the unphysical scale dependence that plague the
prediction of $W+b$-jets cross sections.

The computation of QCD one-loop corrections to $2 \rightarrow n$
processes with $n \geq 4$ is, even in the massless limit, a
challenging task and retaining the mass dependence on internal and
external particles increases the complexity even further.  Therefore,
only a few full NLO cross section computations of $2 \rightarrow 4$
and $2 \rightarrow 5$ processes have been carried out to date, among
them the productions of $W+4 j$ \cite{Berger:2010zx}, $Z + 4j$ \cite{Ita:2011wn},
$4 j$ \cite{Bern:2011ep}, $Z/\gamma + 3 j$ \cite{Berger:2010vm}, $W+3 j$
\cite{Berger:2009zg,Berger:2009ep,Ellis:2008qc,KeithEllis:2009bu,Melnikov:2009wh},
$W + 2 \gamma + j$ \cite{Campanario:2011ud}, $t \overline t j j $
\cite{Bevilacqua:2010ve,Bevilacqua:2011aa}, $t\overline t b\overline
b$ \cite{Bredenstein:2008zb,Bredenstein:2009aj,Bredenstein:2010rs,
Bevilacqua:2009zn}, $b \overline b b \overline
b$~\cite{Binoth:2009rv}, $ W^+W^\pm
jj$~\cite{Melia:2010bm,Melia:2011dw} and $W^+W^-b\overline b$
\cite{Denner:2010jp, Bevilacqua:2010qb}.  A particularly difficult
component of this kind of high-multiplicity processes is the
calculation of virtual one-loop QCD corrections.  The prospect of
increasingly accurate measurements at the LHC triggered a lot of
interest in the improvement and automatization of NLO cross section
predictions.  Primarily two different strategies have been developed
for the evaluation of one-loop corrections: the traditional
Feynman-diagram-based approach as well as unitarity techniques
\cite{Bern:2007dw,Ellis:2011cr}.  Powerful packages like
\texttt{BlackHat} \cite{Berger:2008sj, Berger:2009zg},
\texttt{CutTools} \cite{Ossola:2007ax}, \texttt{Helac-nlo}
\cite{HelacNLO}, \texttt{Rocket} \cite{Giele:2008bc, Ellis:2008qc} and
\texttt{MadLoop}~\cite{Hirschi:2011pa} exist that provide automatization and
efficient numerical implementations of unitarity methods and that
have been successfully applied to the calculation of cutting-edge
one-loop processes.  Recently, the automatized package \texttt{GoSam}
\cite{Cullen:2011ac} has been developed and applied to the automatized
computation of a wide range of NLO cross sections.  Moreover, several
fast and efficient private codes exist that follow the traditional
approach of Feynman diagrams and tensor-integral reduction.

In this work, we develop and describe a new automatized approach to
one-loop calculations based on Feynman diagrams. We test our
techniques against several $2 \rightarrow 3$ and $2\rightarrow 4$
processes for which results are available. For instance, we are able
to reproduce the $\overline ud \rightarrow W d \overline d g$ results
for $W + 3\,j$ production at NLO \cite{Berger:2009ep, Berger:2010vm}.
Finally, we apply them to the novel computation of one-loop
corrections to the cross section of $qg \rightarrow W b \overline b
q'$ with a massive $b$ quark.

The paper is organized as follows: In
section~\ref{sec:general_strategy}, we present the anatomy of the
parton-level processes and discuss the general strategy to generate
and simplify amplitude-specific expressions. Our numerically stable
approach to the evaluation of one-loop tensor integrals is explained
in section~\ref{sec:tensor_reduction}.
Section~\ref{sec:automated_approach} is devoted to the analysis of the
achieved accuracy and computation times. Finally, we conclude by
presenting a numerical result for $\Wbbd$ at NLO for a single
phase-space point in section~\ref{sec:results}.
Section~\ref{sec:conclusions} contains some brief conclusions.

\section{General strategy}
\label{sec:general_strategy}

\begin{figure}[h]
    \begin{center}
        \includegraphics[width=0.95\textwidth]{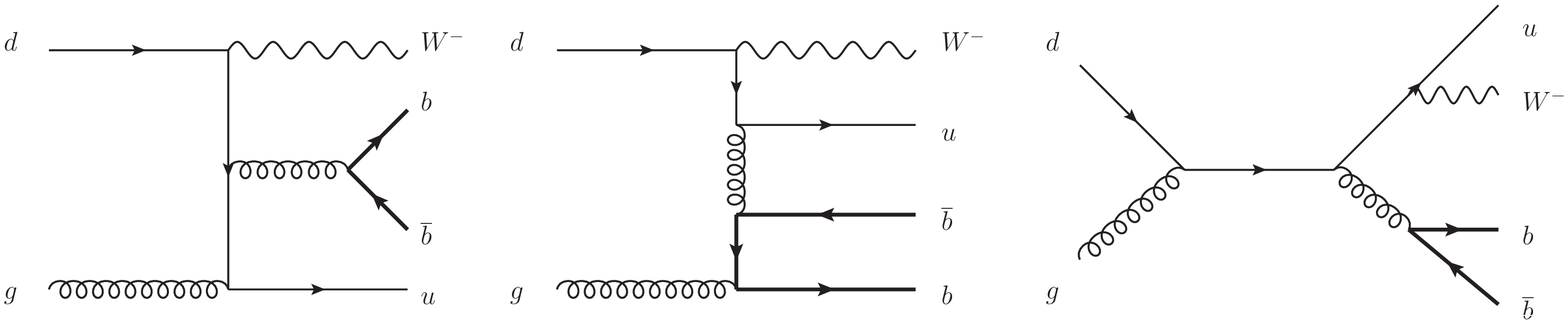}
    \end{center}
    \caption{Example of LO diagrams for $\Wbbd$.}
    \label{fig:LOWbbd}
\end{figure}

At leading order in the strong coupling, the $qg \rightarrow
Wb\overline b q'$ process, with the choice $q = d$ and $q' = u$ which
we consider in the following, consists of 12 tree level diagrams.
Examples of these diagrams are depicted in fig.~\ref{fig:LOWbbd} and
one-loop QCD corrections are obtained by adding virtual gluons and
fermions, yielding 308 Feynman diagrams. Ultra-violet (UV) and
infrared (IR) divergences are regularized with dimensional
regularization in $d = 4-2\epsilon$ dimensions and we keep the full
bottom-quark mass dependence while lighter quarks are treated as
massless. We enforce transversality of external bosons through
$p_W\cdot\epsilon_W = 0$ and $p_g\cdot\epsilon_g = 0$, with
$\epsilon_{W/g}^\mu$ and $p_{W/g}^\mu$ being the polarization vectors
and momenta of the W~boson and gluon, respectively. While this choice
is obvious for gluons, it is justified for the $W$~boson only for weak
couplings to massless fermions, which is the case in the amplitude at
hand.

\begin{figure}[h]
    \begin{center}
        \includegraphics[width=0.55\textwidth]{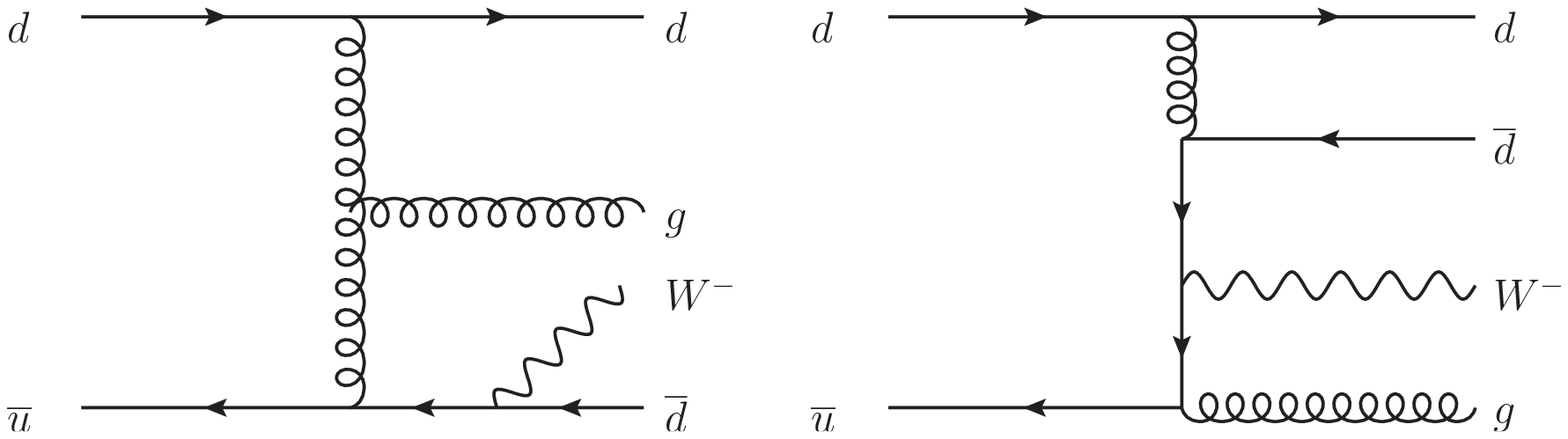}
    \end{center}
    \caption{Examples of additional LO diagrams required for $\Wddg$
      which can not be obtained through crossing of the massless limit
      of $\Wbbd$.}
    \label{fig:LOWddg}
\end{figure}

In the limit of a vanishing bottom-quark mass, the $\Wbbd$ transition
also contributes to $W + 3\,j$ production at NLO. To verify the
correctness of our approach we choose the $W + 3\,j$ subprocess
$\Wddg$, parts of which can be obtained by crossing of the $\Wbbd$
diagrams and setting the bottom-quark mass to zero. Because $d$ quarks
appear both in the initial and final state, however, the number of
diagrams is doubled both at LO and at NLO due to contributions like
the ones depicted in fig.~\ref{fig:LOWddg} and their one-loop
corrections.  We checked our results against Ref.~\cite{Berger:2009ep}
and found agreement.

In the traditional Feynman diagram based approach to the evaluation of
one-loop corrections, a given NLO amplitude $\mathcal M^{(1)}$ is
commonly decomposed as
\begin{equation}
    \mathcal M^{(1)} = \sum_i C_i \sum_{j} c_{ij} I_{j} \hat{\mathcal M}^{(1)}_j
    \label{eqn:NLOampdef}
\end{equation}
with color structures $C_i$ and polarization/spin information
$\hat{\mathcal M}^{(1)}_j$. $I_j$ denotes tensor one-loop integrals
after decomposition into tensor structures of external momenta and
contraction of all Lorentz indices. The sum over $j$ in
eqn.~(\ref{eqn:NLOampdef}) for a given term $C_i$ runs over all
one-loop sub-diagrams with the same color structure.

The color- and spin-summed and/or -averaged squared amplitude is given
by
\begin{equation}
    \Gamma = \text{Re}\left(\sum_{\text{colors}} \sum_{\text{pol}} \mathcal
    M^{(1)} \mathcal M^{(0)*}\right) = \text{Re}\left(\sum_n I_n \Lambda_n\right) 
\end{equation}
with 
\begin{equation}
    \Lambda_n = \sum_{ij} \sum_{colors} C_i C_j^* \sum_{pol} c_{in} \,
\hat{\mathcal {M}}^{(1)}_n \hat{\mathcal {M}}^{(0)*}_j
    \label{eqn:xsectdef}
\end{equation}
where $\mathcal M^{(0)}$ is the leading-order amplitude 
\begin{equation}
    \mathcal M^{(0)} = \sum_i C_i \hat{\mathcal{M}}^{(0)}_i,
\end{equation}

decomposed in color space on the same basis of color structures $C_i$.

After organizing the NLO amplitude by color factor, standard $SU(3)$
relations are applied to simplify the color structures and the
resulting set of color coefficients is extracted.  Summing/averaging
over final/initial color indices, after contraction with the LO color
components, yielding the $C_i C_j^*$ term in
eqn.~(\ref{eqn:xsectdef}), is performed at this point.  In the next
step, tensor integrals are decomposed into Lorentz invariant tensor
coefficients and a standard ordering of Dirac and spinor structures is
achieved with the help of anti-commutation relations of Dirac matrices
and the application of the equations of motion. The amplitude is
subsequently expanded in $(d-4)$ and UV/IR divergences are separated
such that four-dimensional identities can be safely used without
introducing the need for rational terms of either IR or UV origin
\cite{Bredenstein:2008zb}.  Moreover, this approach also avoids
ambiguities in the definition of the $\gamma_5$ matrix, which we treat
in naive dimensional regularization. The complete polarization
information of the amplitude is contained in Dirac chains and
polarization vectors of external bosons, commonly called standard
matrix elements (SME), $\hat{\mathcal M_k}$.  At this stage, the
number of SME is of the order of several thousand for both $2
\rightarrow 4$ processes.  Reducing the set of SME to linear
combinations in a smaller basis $\{\tilde{\mathcal M_k}\}$ is crucial
since the size of final expressions, and therefore the computational
complexity, scales with the number of SME. Algebraic relations based
on four-dimensional identities tailored for the specific process and
SME at hand have been described in
\cite{Denner:2005es,Denner:2005fg,Bredenstein:2008zb,Bredenstein:2009aj,Bredenstein:2010rs}
and successfully applied in several calculations. However, a reduction
to a sufficiently small basis is not straightforward, requires careful
inspection of the individual contributions on a case by case basis,
and is also dependent on the order of the application of different
relations. To automatize this procedure, we have developed a graph
based approach to the SME reduction that allows for an efficient
implementation and performs a brute-force search for a small SME
basis. Products of Dirac chains are translated to directed graphs
where the various structures, like gamma matrices, projection
operators, and spinors are represented by nodes, and directed edges
describe contractions of Lorentz indices and the ordering of
structures.  Algebraic relations then translate to operations on
graphs, for instance shrinking of edges, exchanging or adding of
nodes, and result in general in disconnected graphs.  Since this
method can be expressed very efficiently within the framework of graph
theory without the need for computationally expensive algebraic
manipulations of lengthy expressions, our implementation is capable of
testing a huge number of combinations of transformations. Typically,
the original set of SME is reduced to a basis of several hundred
elements this way. Our variant of the SME reduction will be discussed
in more detail elsewhere~\cite{wbbj_comp}.

As a last step, the products $\mathcal{\hat M}^{(1)}_n \mathcal{\hat{
    M}}^{(0)*}_k$ of the NLO SME with the leading-order color
amplitudes are evaluated, Dirac chains properly contracted, and
summations over spins and polarizations as well as traces are
performed. We translate the resulting expressions into \texttt{C}~code
for an efficient numerical evaluation.

Altogether, the structure of the final evaluation routines is as
follows: each NLO diagram/color amplitude contracted with the tree
level diagrams is expressed through linear combinations of tensor
integral coefficients, products of SME, and kinematic invariants.
The evaluation of tensor coefficients is done as presented in sec.~3,
while products of SME are computed once per phase-space point and
reused. Finally, the sum over squared color amplitudes is evaluated.

It is important to note that, after specifying the desired process and
kinematics, no user interaction is required from the point of diagram
generation to the final numerical code for the cross section
evaluation at single phase-space points. All algebraic manipulations
are performed using \texttt{FORM} while other components like SME and
tensor reductions are developed in \texttt{C++}. Transparent
interfaces, using \texttt{Python}, process input and output between
the different stages and allow for extensive intermediate checks.  The
final cross section evaluation is made accessible through an
automatically generated and flexible \texttt{C++} interface that
allows, for instance, the evaluation of single diagrams or color
amplitudes interfered with the LO contributions, the extraction of
divergences, different reduction methods or a direct connection with a
phase-space generator.

\section{Reduction of tensor integrals}
\label{sec:tensor_reduction}

The integration over the loop momentum in NLO one-loop amplitudes
involves N-point tensor integrals $T^N_{\mu_1\mu_2...\mu_p}$ that are
commonly decomposed into linear combinations of tensor structures
(products of external momenta and metric tensors) with Lorentz
invariant so-called tensor coefficients $T^N_{\{j\}}$. The indices
$\{j\}$ encode the rank and composition of the corresponding tensor
structure.  The general strategy for the evaluation of $T^N_{\{j\}}$
is their reduction to master integrals, usually scalar $N$-point
functions $T^N_0$.  In the case of $N$-point tensors with $N \leq 4$ the well-known
Passarino-Veltman (PV) algorithm~\cite{Passarino:1978jh} can be used,
while $N>4$ coefficients are reduced to linear combinations of
four-point tensor integrals. Due to numerical instabilities in the
vicinity of phase-space points where Gram determinants become small,
alternative reduction techniques exist to produce reliable results.

\paragraph{Applied methods} 

Our tensor reduction approach combines different methods, allows for
cross checks between them and ensures numerical stability in an
automatized way.  For $N$-point functions with $N \leq 4$ the
following reduction schemes are used: 
\begin{itemize} 
\item PV reduction \cite{Passarino:1978jh}, 
\item reduction with modified Cayley determinants as introduced by
  Denner and Dittmaier (DD) in \cite{Denner:2005nn}, and 
\item expansions around small quantities, like Gram/Cayley
  determinants and kinematic invariants (DDx) developed by the same
  authors of Ref.~\cite{Denner:2005nn}.  
\end{itemize} 
In addition, our software is capable of producing multiple precision
(MP) reductions with help of the \texttt{qd} library \cite{libqd},
that turn out to be numerically stable already in the framework of the
PV reduction (MP~PV).  Tensor coefficients with five and six external
legs are evaluated following an approach by Diakonidis et al.
\cite{Diakonidis:2008ij, Diakonidis:2008dt} that is free of inverse
Gram determinants (GDF) and therefore numerically stable.

Our implementation is inspired by \cite{Denner:2005nn} and performs
the reduction numerically.  However, the original recursive algorithm
is unrolled into an iterative procedure by arranging the tensor
coefficients in a tree-like structure, which provides fine-grained
control over different aspects of the reduction.  We choose the PV
reduction for $N\leq4$ in the absence of numerical instabilities and
the GDF reduction for $N>4$ as our standard methods.  Based on these
reductions, the evaluation tree is constructed for the required set of
tensor coefficients $T^N_{\{j\}}$ in such a way that the minimal
number of evaluations is guaranteed. For optimal reuse of intermediate
results, coefficients with different mass distributions on internal
propagators are brought to a standard form with respect to the
external momenta and internal masses and are treated together. As an
example consider a completely massless 3-point function coefficient
that is required for some subdiagrams in the amplitude. The same
coefficient may appear as a dependency in the reduction of several
4-point functions, which can be either massless or contain one massive
propagator. The nodes of the evaluation tree are assigned the default
reduction strategy and after initial creation the tree is reused for
the majority of phase-space points. As already mentioned, this
strategy works well in large regions of phase space, but becomes
numerically unstable if small Gram determinants in $N\leq4$ point
coefficients are encountered. In this case, the evaluation tree is
extended with subtrees for the unstable tensor functions and their
dependencies only. These newly created subtrees are computed with one
of the alternative methods, either DD or DDx as needed, or with MP PV,
to ensure numerical stability, while all other nodes are reduced with
the default procedures.  While both approaches provide numerically
stable results, we use the former techniques mainly for cross checks
in critical phase-space regions while we employ the latter in
computations of squared amplitudes.  As subtrees are added, the
reduction program keeps track of the conditions that lead to
inconsistencies such that the newly created evaluation paths can be
reused in future evaluations.

\paragraph{Numerical stability}

Detecting numerical instabilities at a single phase-space point is in
general a non-trivial task without examining the surrounding
phase-space domain or additional external information. Performing the
tensor reduction in $d = 4 - 2\,\epsilon$ dimensions and regulating
both ultraviolet and infrared divergences dimensionally, however,
offers a direct handle on the achieved accuracy. Firstly, the scalar
one-loop integrals in terms of which the tensor-integral coefficients
are reduced have to be known retaining the full pole structure. For
this task, we use a custom implementation based on \texttt{QCDLoop}
\cite{Ellis:2007qk} for the IR poles together with a modified version
of \texttt{LoopTools} \cite{Hahn:1998yk} that allows for multiple
precision evaluations.  During the reduction which is performed on the
divergent and finite parts separately, UV/IR poles are affected by the
same numerical instabilities as the finite part.  Provided the
divergences can be computed for a given tensor coefficient
independently in a reliable way, a direct comparison can be used to
detect a loss of precision\footnote{At the cross section level, the
same approach to identify numerical instabilities has been
successfully used in \cite{Berger:2008sj,Berger:2008sz,
Berger:2009ep}. }. UV poles of tensor coefficients in the minimal
subtraction scheme are mass independent and can either be obtained
with a single analytic reduction or can be taken from the appendix of
\cite{Denner:2005nn}.  In contrast, IR divergences occur in certain
limits of vanishing kinematic invariants and have to be studied in all
these cases individually. To this end, we have reduced 3- and 4-point
tensor coefficients up to rank~4 analytically in the various IR
divergent limits.  Using these results a numerically stable library
for their computation was carefully established and thoroughly
cross-checked with the full analytical results for a wide range of
input parameters.  In our reduction, the UV and IR poles of each
tensor coefficient are then checked against this library for every
phase-pace point and, in case of inconsistencies, alternative
reduction methods are employed as described above. Of course, this
procedure comes with a computational cost, that is mainly due to the
multiple reductions for pole and finite parts and the evaluation of
scalar integrals, while the contribution from the IR pole evaluation
routines are negligible.  However, thanks to the efficient design of
the reduction algorithm and extensive caching, the run times are
competitive with what is reported in the literature: computation times
in the numerically stable case for all tensor integrals required in a
mixed massive and massless $2\rightarrow 4$ process average at around
20~ms per phase-space point on an Intel i7 950 CPU at 3.07GHz.

\section{Characteristics of the automatized approach}
\label{sec:automated_approach}

In its current state, our automatized approach is tested for QCD NLO
one-loop corrections to amplitudes with up to six external particles,
of which at least one and at most three are (weak or strong) gauge
bosons. We have cross checked several $2 \rightarrow 3$ and $2
\rightarrow 4$ processes with the literature. For instance we have
reproduced full results for $Wb\overline b$/ $Zb\overline b$
hadroproduction~\cite{Febres Cordero:2006sj, Cordero:2007ce,
Cordero:2009kv} and we have checked individual parts of the NLO
calculation of hard-photon production with heavy quarks ($Q\overline Q
\gamma$ for $Q=b,t$) against an independent internal
calculation~\cite{LRTSHH}. The $2 \rightarrow 4$ process $\overline u
d \rightarrow W d \overline d g$ was checked at one benchmark point
against a result obtained in \cite{Berger:2009ep}.

Apart from instabilities in the reduction of tensor coefficients,
cancellations in intermediate expressions of the unrenormalized
squared amplitude $\Gamma$ in eqn.~(\ref{eqn:xsectdef}) may also
induce a loss of accuracy in some phase-space regions.  In this case,
we extend the numerical precision for both the complete tensor
reduction as well as the evaluation of the whole contribution to
$\Gamma$.  Again, the loss of precision is detected by comparison with
the expected UV divergence structure of $\Gamma$, which is independent
of the used renormalization scheme and can be obtained with high
precision from counterterm contributions. This step is computationally
most expensive, as a huge number of operations has to be performed in
slow multiple precision mode both in the tensor reduction and in the
evaluation of $\Gamma$.  Fortunately, the proportion of this type of
evaluations is in general relatively small.  Compared to the naive
approach where no analysis of instabilities is performed on the tensor
reduction level, the necessary number of this kind of evaluations is
substantially reduced.

Table \ref{tab:benchmarks} gives an overview of the obtained
efficiency for the evaluation at $5\cdot10^4$ random phase-space
points with reasonable cuts, requesting a maximal relative error of
$10^{-5}$.  As expected, the evaluation time scales with the number of
external particles.  Moreover, due to a larger basis of SME,
amplitudes containing weak couplings compared to for example $t
\overline t \gamma$ production are computationally more expensive.  An
interesting observation, however, is the fact that the number of
switches to multiple precision evaluations, both within the reduction
and at the amplitude squared level, do not vary much between processes
of comparable complexity. Although evaluations in quadruple precision
take significantly more time with increasing number of external
states, the overall evaluation time is governed by the numerically
stable bulk of phase space.

\begin{table}
\begin{center}
\begin{tabular}{l|rrrrrrrr}
Process & $r_s$ & $r_q$ & $r_{dq}$ & $t_m$/ms & $t_s$/ms & $t_q$/ms & $t_{dq}$/ms & $t_{q}^{\text{full}}$/ms \\\hline\hline
$q\overline q \rightarrow \gamma t \overline t$ & 99.6\% &   0.4\% &   0 &     9.5  &  8.9  &  153  & 0     &  1069 \\
$g g \rightarrow \gamma t \overline t$          & 98.9\% &   1.1\% &   0 &     12.0 &  10.1 &  182  & 0     &  1972 \\    
$q\overline q' \rightarrow W b \overline b$      & 99.7\% &   0.3\% &   0 &     10.9 &  10.4 &  167  & 0     &  1264 \\    
$q\overline q \rightarrow Z b \overline b$      & 99.8\% &  0.1\% &   0.1\%& 17.7 &  14.4 &  217  & 3161  &  2290 \\    
$gg \rightarrow Z b \overline b$                & 98.3\% &  1.6\% &   0.1\%& 22.5 &  15.7 &  233  & 3314  &  2706 \\    
$\overline u d \rightarrow d \overline d g W$   & 95.4\% &  3.6\% &   1.0\%&   90.3 &   37.5&  306  & 4358  &  5503\\   
$ u g \rightarrow b \overline b d W$            & 93.1\% &  5.6\% &   1.3\%&   95.4 & 29.7  &  311  & 3870  &  5192
\end{tabular}

\caption{Benchmarks of the numerically stabilized method applied to
  various NLO amplitudes for the evaluation of $5\cdot10^4$
  phase-space points.  $r_s$, $r_q$ and $r_{dq}$ give the ratios of
  phase-space points that required either only standard (double) or
  also some additional quadruple/double-quadruple precision
  evaluations at the reduction or amplitude-squared level for reliable
  numerical results. $t_m$ gives the mean evaluation time per
  phase-space point while $t_s$, $t_q$ and $t_{dq}$ denote separate
  mean timings for the respective numerical precision.  Finally, the
  mean computation time of both the amplitude and tensor reduction in
  full quadruple precision is given in $t_{q}^{\text{full}}$.  The
  above numbers were obtained on an Intel i7 950 CPU at 3.07GHz.}
\label{tab:benchmarks}
\end{center}
\end{table}

\begin{figure}
    \begin{center}
        \includegraphics[scale=0.9]{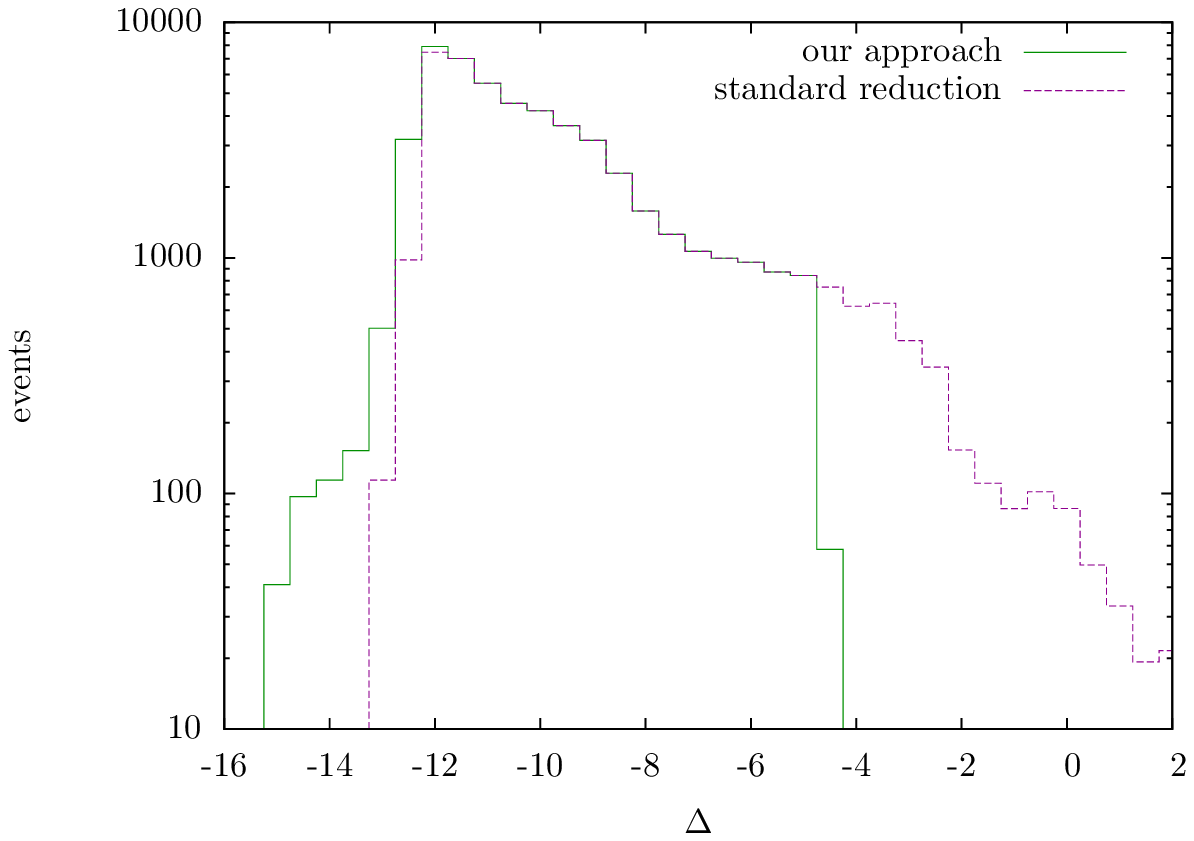}
    \end{center}
    \caption{Comparison of the obtained numerical precision for the
      finite part of the process $dg \rightarrow Wb \overline b u$ at
      $5\times 10^4$ phase-space points using standard double precision
      PV tensor reductions and our numerically stable approach.}
    \label{fig:Accuracy}
\end{figure}

It is instructive to study the effect of both our approach and
different choices of reduction algorithms on the obtained accuracy in
the final result. First, the unrenormalized squared NLO amplitude of
$\Wbbd$ is sampled in multiple precision for $5\cdot 10^4$ different
phase-space points and the UV divergent parts are verified to cancel
with counterterm contributions. The so-obtained results serve as
high-precision reference points $\Gamma_{\text{ref},i}$. 
Subsequently, we compute the squared
amplitude for the same set of points using two different strategies:
\begin{enumerate}
    \item standard reduction of 5- and 6-point tensor coefficients,
    \item GDF reduction of 5- and 6-point tensor coefficients with
          switches to multiple precision both at the $n$-point tensor
          integral reduction and
          amplitude squared level when required, requesting a maximal
          relative error of $10^{-5}$
\end{enumerate}
and $N\leq4$-point functions are treated with the PV reduction
algorithm.  Fig.~\ref{fig:Accuracy} shows the distribution of the
logarithmic error,
\begin{equation}
   \Delta_i = \log_{10}\left(\frac{|\Gamma_i - \Gamma_{\text{ref},i}|}
{|\Gamma_{\text{ref}, i}|}\right),
\end{equation}
of the squared amplitude $\Gamma$ relative to the reference points
$\Gamma_\text{ref}$ for the two different cases.  Without special
handling of numerical instabilities, strategy~1, as expected shows a
wide distribution of the obtained relative error. Our approach,
strategy~2, gives a considerably better relative error distribution by
reevaluating low precision points.

\section{Result for $\Wbbd$} 
\label{sec:results}

For future reference, we provide our new result for the unrenormalized
squared amplitude of $\Wbbd$ at NLO at a single phase-space point. The
result is normalized to the LO cross section in the following way
\begin{equation}
   \hat \Gamma = \frac{(4\pi)^{2- \epsilon} }{8\pi\alpha_s} 
\frac{\Gamma(1-2\epsilon)}{\Gamma(1+\epsilon)\Gamma^2(1-\epsilon)} 
\frac{\Gamma}{\left|\mathcal M^{(0)}\right|^2},
\end{equation}
such that the final result is independent of the strong and weak
couplings as well as CKM matrix elements. Furthermore, we use
\begin{eqnarray}
    m_W &=& 80.41 \, \text{GeV and} \\
    m_b &=& 4.62 \, \text{GeV}.
\end{eqnarray}
for the weak-boson and bottom-quark masses and set all external
particles on-shell. 

For $\Wbbd$ our result with $n_l = 4$ light and $n_h = 1$ heavy-quark
flavors at the phase-space point of tab.~\ref{tab:ps1} with
renormalization scale $\mu^2 = (p_d + p_g)^2$ reads

\begin{equation}
    \hat\Gamma(\Wbbd) = -5.6666667\;\epsilon^{-2} + 39.342424 \;\epsilon^{-1} + 292.92493
\end{equation}

\begin{table}
\begin{center}
\footnotesize{
\begin{tabular}{c|r|r|r|r}
          &            $E$ & $p^1$ & $p^2$ & $p^3$ \\\hline\hline
    $p_d$ &           $100.000000000000 $ & $                 0 $ & $                 0 $ & $  100.000000000000     $ \\
    $p_g$ &           $100.000000000000 $ & $                 0 $ & $                 0 $ & $ -100.000000000000 $ \\
    $p_u$ &           $14.4169546267975 $ & $ -3.59819144566031 $ & $  6.52544251406004 $ & $ -12.3418069595668 $ \\
    $p_b$ & $53.6542637065835 $ & $ -16.9076522158373 $ & $ -49.1575349754512 $ & $  12.4540622120327 $ \\
    $p_{\overline b}$ &           $25.2318438952597 $ & $ -17.2383739318242 $ & $ -15.9080092164594 $ & $  8.06692341047065 $ \\
    $p_W$ &           $106.696937771359 $ & $  37.7442175933219 $ & $  58.5401016778506 $ & $ -8.17917866293656 $ \\
\end{tabular}
}
\end{center}
\caption{Phase-space point used for $\Wbbd$}
\label{tab:ps1}
\end{table}

\section{Conclusions}
\label{sec:conclusions}
We have developed a new automatized approach to the evaluation of
one-loop amplitudes in terms of Feynman diagrams and applied it to the
calculation of the $\mathcal O(\alpha_s)$ virtual corrections to $qg
\rightarrow W b \overline b q'$ (and $q\overline q^\prime\rightarrow W
b \overline b g$). These corrections enter the NNLO calculation of
$Wb\bar{b}$ hadroproduction as well as the NLO calculation of both $W
b \overline b + j$ and $W b + j$ production in a fully consistent
four-flavor-number scheme. A thorough study of the impact of these
corrections in both previous cases as well as the application of the
method developed in this paper to other processes will be the subject
of future publications.

\section*{Acknowledgments} 
The authors would like to thank Thomas Reiter for providing results
that confirmed the main result of this work.  This work is supported
in part by the U.S. Department of Energy under grant DE-FG02-97IR41022
and by the National Science Foundation under Grant No. NSF
PHY05-51164. L.R. would like to thank the Kavli Institute for
Theoretical Physics (KITP) for the kind hospitality while this work
was being completed.

\end{document}